\documentclass[aps,amsmath,preprint,prb,floats,tightenlines]{revtex4}
\usepackage{graphicx}
\newcommand{\be}{\begin{eqnarray}}
\newcommand{\ee}{\end{eqnarray}}
\newcommand{\nn}{~\nonumber \\}

\newcommand{\figu}[2]{
 \begin{center}
  \begin{figure}
   \includegraphics{#1_punkt}
   \vspace{5mm}
   \caption{#2}
   \label{#1}
  \end{figure}
 \end{center}
}

\begin{document}
\title{Quasi-M\"ossbauer effect in two dimensions}
\author{Dennis D. Dietrich and Lester L. Hirst\footnote{deceased}}
\affiliation{
Institut f\"ur Theoretische Physik, 
Johann Wolfgang Goethe-Universit\"at, 
60054 Frankfurt am Main,
Germany}
\begin{abstract}
Expressions for the absorption spectrum of a nucleus in a three- and a
two-dimensional crystal respectively are obtained analytically at zero and 
at finite temperature respectively. It is found that for finite temperature in 
two dimensions the 
M\"ossbauer effect vanishes but is replaced by what we call a
Quasi-M\"ossbauer effect. Possibilities to identify two-dimensional elastic
behavior are discussed.
\end{abstract}
\maketitle
\section{Introduction}

In nuclear reactions, the atoms usually are treated as free, i.e., unbound. 
The chemical bonds between each other are neglected. This is justified with 
the argument that the typical energy in a nuclear process is generally much 
higher than the chemical binding energies ($1$ to $10$ eV) and moreso than 
that of lattice vibrations ($10^{-2}$ to $10^{-1}$ eV) \cite{wertheim}. In 
return, most of the detailed qualities of the nucleus are of secondary 
importance in solid state physics.

These assumptions are not correct without limitations. Consider for example 
the absorption or emission of $\gamma$-photons by nuclei. During absorption
processes (emission analogously) the nucleus first is in a ground state and 
makes a transistion into an excited state. The energy difference between the 
ground and the excited state is not constant. It has a certain distribution 
which is linked to the life-time of the excited state by Heisenberg's 
uncertainty relation. For these reactions the difference has values between $10^{-11}$
to $10^{-6}$ eV \cite{wertheim}. Thus, it is considerably smaller than the 
energy of chemical bonds.

During the absortion event, the atom would recoil due to the conservation of
the sums of the momenta of the atom before the reaction and the absorbed 
photon. The thermal movement of an atom in a gaseous absorber leads to a 
Doppler broadening of the absorption line and the recoil to a shift of its
maximum. Typical 
recoil-energies lie in the range from $10^{-5}$ to $10^{+2}$ eV 
\cite{wertheim}. This interval also comprises the energy of chemical bonds, 
phonons, and even energetically lower areas.

For atoms bound in a system and photon momenta below the threshold of the 
destruction of the chemical bonds, there are different aspects coming into 
play. For molecular systems additional rotational and vibrational degrees of 
freedom can be important. Simultaneously, the Doppler broadening and the 
recoil of the system decrease due to the enhanced mass of the system.

Making the transition to a crystal, the Doppler broadening and the recoil
of the entire system become negligible. The entire recoil-energy goes into 
lattice vibrations. Further, there is the possibility that the momentum of
the photon is taken up by the entire crystal without producing any phonons. 
Such an event is called {\it recoil-free}. In a rigid crystal, the absorption
spectrum would just show the pure absorption spectrum of the nucleus, 
because no phonons can be produced. For photons with very low energy, i.e., 
less than $1.5\times10^{+5}$ eV the recoil-energy is below the energy of the 
phonons. Classically, no vibrational modes of the crystal could be excited, 
thus the energy and momentum conservation could not be satisfied and the 
related processes could not take place. However, observations show that they
do take place and that in those situations a part of the events are recoil-free. 
Energy and momentum conservation are guaranteed when averaging 
over many events \cite{lipkin}. The effect and the fraction of recoil-free
events have been named after its discoverer R.~L.~M\"ossbauer 
\cite{moessbauer}.

The above discussion also plays a r\^ole in the scattering of x-rays
or neutrons from atoms. There, the distinction between elastic and 
inelastic events is also known. M\"ossbauer used also results by Lamb about 
neutron scattering \cite{lamb}. This is the reason for the name 
M\"ossbauer-Lamb fraction.

Of interest is not only the total intensity not contained in the M\"ossbauer
peak but also its distribution onto the remaining spectrum. Here, Visscher 
\cite{visscher} derived an analytical result for the observable spectrum at 
zero temperature in a three dimensional crystal. He also performed numerical
calculations for finite temperatures. Systems with almost two-dimensional
magnetic interactions are known (interaction strength $10^4$-times stronger in
the plane than between the planes). Now, it is interesting to identify 
systems with low-dimensional elastic interactions. Here, one could think 
about ultrathin films sputtered onto the surface of a solid, a liquid, 
or the boundary layer between two liquids. Thus a mechanical decoupling to the 
substrate should be achieveable at least in directions parallel to the 
surface.

In the following section we discuss the theoretical description of these
phenomena by deriving and solving a differential equation for the
spectrum. Subsequently, in section III, we evaluate the solution in harmonic
approximation for the isotropic Debye model in three and two dimensions
at zero and finite temperature. In two dimensions at finite temperature we
find that there are no recoil-free events. The M\"ossbauer effect ceases to
exist in the strict sense. However there is still a central peak made up of
almost recoil-free events, a Quasi-M\"ossbauer effect. Ways how to identify
two dimensional elastic behavior are discussed. In the final section we
summarise our paper.

\section{Theory}

Starting from the absorption spectrum of a nucleus inside a rigid crystal 
lattice $S_{rig}(\omega_\gamma)$, we write the observed spectrum in an elastic 
lattice $S(\omega_\gamma)$ as a convolution of the first with a shift 
spectrum: $S=I*S_{rig}$. Here, $\omega_\gamma$ is the circular frequency 
of the converted photon.
The shift spectrum $I(\omega)$ comprises all the information on the 
modifications of the observed spectrum $S(\omega_\gamma)$ due to the 
dynamics of the lattice. Up to an overall factor $I(\omega)$ is the
spectrum one would observe, if the nucleus' line width was exactly zero.
A normalisation to one $\int d\omega I(\omega)=1$
guarantees the conservation of the integrated intensity: 
$\int d\omega S(\omega)=\int d\omega S_{rig}(\omega)$.

In a crystal with $N$ Bravais lattice sites, $A$ atoms per elementary cell, 
and where the atoms are elongated in $d'$ dimensions there are $M=NAd'$ 
vibrational modes with circular frequencies $\Omega_J$ with 
$J\in\{1,2,...,M\}$.
In general, the dimensionality of the crystal $d$ is the same as that of the 
elongations of the atoms $d'$ which does not exclude the possibility of doing 
calculations with $d\neq d'$.

The state of the crystal before ($_i$) and after ($_f$) the emission or 
absorption process can be described by a set of $M$ numbers each 
$\{n\}^{i,f}=\{n_J^{i,f}:J\in\{1,2,...,M\}\}$.
The set of all possible changes of the state is constructed by combining 
every possible initial state with every possible final state 
(direct product).
The different changes contribute to the shift spectrum $I(\omega)$ with 
different weights $A(\{n\}^i\rightarrow\{n\}^f)$. The shift
spectrum can be expressed as the weighted mean over all changes conserving 
the total energy:

\be
I(\omega)
=
\sum_{\{n\}^i}\sum_{\{n\}^f}
\delta[\hbar\omega-(E^f-E^i)]
A(\{n\}^i\rightarrow\{n\}^f).
\label{discreteshift}
\ee

In a harmonic crystal the total energy is given by: 

\be
E^{i,f}=\hbar\sum_J\left(n_J^{i,f}+\frac{1}{2}\right)\Omega_J.
\label{harmonicenergy}
\ee

Due to the linearity of the above expression, the energy difference in 
equation (\ref{discreteshift}) is determined exclusively by the change

\be 
m_J={n_J^f}-{n_J^i}
\label{change}
\ee

of the 
occupational numbers in the different modes but not from the state 
$\{n\}^i$  of the crystal before the absorption or emission process.
The weight $A(\{n\}^i\rightarrow\{n\}^f)$ is equal to the absolute
probability for the transition from the state $\{n\}^i$ to the state
$\{n\}^f$. It is given by the 
product of the probability $A(\{n\}^i)$ of the crystal to be in a 
state $\{n\}^i$ before the process and the conditional 
probability $A_{\{n\}^i}(\{n\}^f)$ to reach a final state 
$\{n\}^f$ coming from an initial state $\{n\}^i$:

\be
A(\{n\}^i\rightarrow\{n\}^f)
=
A(\{n\}^i)
A_{\{n\}^i}(\{n\}^f).
\label{weight}
\ee

Substitution of equations (\ref{harmonicenergy}) and (\ref{weight}) into 
equation (\ref{discreteshift}) leads to:

\be
I(\omega)
&=&
\sum_{\{m\}}
\delta\left(\hbar\omega-\hbar\sum_Jm_J\Omega_J\right)
\sum_{\{n\}^i}
A(\{n\}^i)
A_{\{n\}^i}(\{n\}^f),
\ee

where the elements of $\{n\}^f$ are to be replaced in accordance with equation
(\ref{change}).
For a harmonic crystal in thermal equilibrium before the elementary 
process, the terms belonging to a certain mode are independent from the 
configuration of the rest of the system. After defining the thermally averaged 
-- i.e., averaged with the thermal weight $P_{n_J^i}^J$ for having $n_J^i$
phonons in mode $J$ before the process \footnote{In thermal equilibrium:
$A(\{n\}_i)=\prod_JP_{n_J^i}^J$.} -- 
probability for the production of $m_J$ phonons in mode $J$:

\be
A_{m_J}^J
=
\sum_{_in_J}
P_{_in_J}^J
~
A_{_in_J\rightarrow_in_J+m_J}^J,
\ee

where $A_{_in_J\rightarrow_in_J+m_J}^J$ is the probability for a change of 
the phonon number in mode $J$ from $_in_J$ to $_in_J+m_J$ \footnote{The probability for the change of two configurations of the
crystal into another is given by the product of the probabilities for the
respective changes in all the modes: 
$A_{\{n\}^i}(\{n\}^f)=\prod_J~{A_{{n_J^i}\rightarrow n_J^i+m_J}^J}$.},
the shift spectrum can be reexpressed as:

\be
I(\omega)
&=&
\sum_{\{m_J\}}
\delta\left(\hbar\omega-\hbar\sum_Jm_J\Omega_J\right)
\prod_{J'}A_{m_{J'}}^{J'}.
\label{thermalshift}
\ee

The probabilities are normalised to one: $\sum_{m_J}A_{m_{J}}^{J}=1$.
Consider the shift spectrum $I^{(J)}(\omega)$ for only 
$J$ active modes. The shift spectrum $I^{(0)}(\omega)$ for a nucleus in an 
entirely rigid crystal lattice is given by a Dirac-$\delta$-distribution: 
$\delta(\omega)$. The shift spectrum $I^{(M)}(\omega)$ with all modes in 
action is identical to the actual shift spectrum $I(\omega)$. From equation 
(\ref{thermalshift}) we get a recursion equation for the partial shift spectra:

\be
I^{(J)}(\omega)=\sum_{m_J}A_{m_J}^JI^{(J-1)}(\omega-m_J\Omega_J).
\ee

An analogous equation is valid for the changes 
$\Delta I^{(J)}=I^{(J)}(\omega)-I^{(J-1)}(\omega)$ in connection with the 
shift amplitudes $a_{m_{J}}^{J}=A_{m_{J}}^{J}-\delta_{m_{J},0}$:

\be
\Delta I^{(J)}(\omega)=\sum_{m_J}a_{m_J}^JI^{(J-1)}(\omega-m_J\Omega_J).
\label{differentialrecursion}
\ee

The shift amplitudes $a^J_{m_J}$ differ from the probabilities
$A_{m_J}^J$ only for $m_J=0$. There the respective shift amplitude is
smaller by one. For example $A_{m_J}^J=0$ for all $m_J\neq 0$ and $A_0^J=1$
signifies that no phonons are created or annihilated in mode $J$; so does
$a_{m_J}^J=0$ for all $m_J$. Hence $A_0^J$ is a meassure for remaining in
the previous state and $a_0^J$ indicates the relative change. Because of the 
additional Kronecker tensor $\delta_{i,j}$ and the 
normalisation of the probabilities $A^J_{m_J}$, the sum over all shift 
amplitudes amounts to zero: $\sum_{m_J}a_{m_{J}}^{J}=0$. Writing the 
shift spectrum $I^{(J)}(\omega)$ as a sum over all its changes and replacing
the changes $\Delta I^{(J)}(\omega)$ of the shift spectrum by the recursion
(\ref{differentialrecursion}) yields:

\be
I^{(J)}(\omega)
=
I^{(0)}(\omega)
+
\sum_{J'=1}^J\sum_{m_{J'}}a_{m_{J'}}^{J'}I^{(J'-1)}(\omega-m_{J'}\Omega_{J'}).
\label{differentialshiftspectrum}
\ee

In case, several active branches are present, the label of the mode $J$ is 
composed of a branch
index $j$ and a Bloch vector $\vec k$: $J=(j,\vec k)$. In a transition from 
a discrete to a continuous description the summation over all modes $J$ is 
replaced by a summation over the different modal branches and an integration 
over $k$-space. The $d$-dimensional integral is substituted by a 
one-dimensional 
integration over the frequencies $\Omega$ of the modes. As a correction, the
density of modes $\eta_j(\Omega)$ for the corresponding branch has to be 
taken into account.
In the following, all investigations will be carried out for a single active 
branch. So, with the volume $V$ of the crystal:

\be
\sum_{J}
\rightarrow
V\int\frac{d^dk}{(2\pi)^d}
\rightarrow
\int d\Omega\eta(\Omega).
\ee

With the dispersion relation $\Omega(\vec k)$ the density could be given by 
\cite{ashcroftmermin}:

\be
\eta(\Omega)
=
V\int\delta[\Omega-\Omega(\vec k)]\frac{d^dk}{(2\pi)^d}.
\label{densityofmodes}
\ee

It can be interpreted as the total number of modes in the infinitessimally 
small interval $[\Omega,\Omega+d\Omega]$. The modes are counted in the 
entire crystal and not only in a unit volume. So, the density of modes
grows proportionally to the number $N$ of Bravais lattice sites. It is to 
vanish for frequencies outside the intervall $[0,\Omega_M]$. When connected 
to the discrete description this must also be the interval in which lie all the 
modal frequencies $\Omega_{J}$ with $J\in\{1,2,...,M\}$. In the Debye 
model, $\Omega_M$ is equal to the Debye frequency $\Omega_D$. For simple models, the 
alternative to marking the shift spectrum $I^{(J)}(\omega)$ and the shift
amplitudes $a^J_{m_J}$ with the mode index $J$ is to express them as 
functions of the modal frequency $\Omega_J$: 
$I^{(J)}(\omega)=I(\Omega_J,\omega)$ and $a^J_{m_J}=a_m(\Omega_J)$. During the 
transition from the discrete to the continuous description the discrete 
parameter $\Omega_J$ is replaced by the continuous $\Omega$. Values at 
$\Omega=\Omega_J$ for all $J\in\{1,2,...,M\}$ are preserved:
$\Omega_J,~J\in\{1,2,...,M\}\rightarrow\Omega\in[0,\Omega]$,
$I(\Omega_J,\omega)\rightarrow I(\Omega,\omega)$,
and
$a_m(\Omega_J)\rightarrow a_m(\Omega)$. Carrying out the continuum limit in 
equation (\ref{differentialshiftspectrum}) leads to:

\be
I(\Omega,\omega)
=
\delta(\omega)
+
\int_0^\Omega d\Omega'
\eta(\Omega')\sum_ma_m(\Omega')I(\Omega',\omega-m\Omega').
\label{continuum}
\ee

Here -- for the sake of simplicity -- we anticipate a result that will be 
derived later on: The shift amplitudes $a_m(\Omega)$ for fixed argument are 
proportional to integer negative powers of the number $N$ of Bravais lattice 
sites: $a_m(\Omega)\sim N^{-|m|}$ for all $m\neq0$ and 
$a_0(\Omega)\sim N^{-1}$. The shift amplitudes occur only 
combined in a 
product with the density of modes. As already stated above, the density of 
modes is proportional to the number $N$ of Bravais-lattice sites. So, only 
the three terms with $|m|\le 1$ survive taking the continuum limit. Two of the three
shift amplitudes that are left can be expressed as functions of the third:
They can be eliminated with the help of the normalisation of the shift 
amplitudes and because we are assuming thermal equilibrium:
$a_{-m}(\Omega)=e^{-m\beta\hbar\Omega}a_{+m}(\Omega)$, respectively, where 
$\beta=k_BT$. Partial differentiation of 
equation (\ref{continuum}) with respect to the maximum frequency $\Omega$ 
together with the above facts leads to a linear, first-order, non-local, 
partial differential equation:

\be
\frac{\partial}{\partial\Omega}I(\Omega,\omega)
=
\eta(\Omega)a_{+1}(\Omega)
\{
e^{-\beta\hbar\Omega}I(\Omega,\omega+\Omega)
-
(e^{-\beta\hbar\Omega}+1)I(\Omega,\omega)
+
I(\Omega,\omega-\Omega)
\},
\label{pde}
\ee

with the starting condition $I(0,\omega)=\delta(\omega)$.
A Fourier transformation from the variable $\omega$ to the variable $\phi$ and
a subsequent division through the Fourier transformed shift spectrum 
$\tilde I(\Omega,\phi)$ leads to an ordinary first-order differential equation:

\be
\frac{d}{d\Omega}\ln\{\tilde I(\Omega,\phi)\}
=
\eta(\Omega)a_{+1}(\Omega)
\{
e^{-\beta\hbar\Omega}(e^{+i\phi\Omega}-1)
+
(e^{-i\phi\Omega}-1)
\}.
\label{fourierdifferential}
\ee

The actual shift spectrum $I(\omega)$ is equal to 
$I(\Omega=\Omega_M,\omega)$. Taking into account the Fourier transformed 
boundary condition $\tilde I(\phi=0)=1$, we find by direct integration:

\be
\tilde I(\phi)
=
\exp\left\{
\int_0^{\Omega_M}d\Omega\eta(\Omega)a_{+1}(\Omega)
[e^{-\beta\hbar\Omega}(e^{+i\phi\Omega}-1)+(e^{-i\phi\Omega}-1)]
\right\}.
\label{fouriershift}
\ee

Given the existence of the resulting integrals, the integral in the exponent
can be seperated into a $\phi$-dependent and a $\phi$-independent part. A 
Fourier retransformation yields the shift spectrum $I(\omega)$ in momentum 
representation:

\be
I(\omega)
=
e^{-F}
\int\frac{d\phi}{2\pi}
e^{i\phi\omega}
\exp\left\{
\int_0^{\Omega_M}d\Omega\eta(\Omega)a_{+1}(\Omega)
[e^{(+i\phi-\beta\hbar)\Omega}+e^{-i\phi\Omega}]
\right\}.
\label{mlf}
\ee

The factor $e^{-F}$ with
$
F
=
\int_0^{\Omega_M}d\Omega\eta(\Omega)a_{+1}(\Omega)\{e^{-\beta\hbar\Omega}+1\}
$ 
is equal to the M\"ossbauer-Lamb fraction.
The exponential series is uniformly convergent on every bounded area of the 
complex plane. So, as long as the magnitude of the exponent is finite, the 
exponential series can be integrated term by term:

\be
I(\omega)
=
e^{-F}
\sum_{\nu=0}^{\infty}\frac{1}{\nu!}
\int\frac{d\phi}{2\pi}
e^{i\phi\omega}
\left\{
\int_0^{\Omega_M}d\Omega\eta(\Omega)a_{+1}(\Omega)
[e^{(+i\phi-\beta\hbar)\Omega}+e^{-i\phi\Omega}]
\right\}^\nu.
\ee

Reinterpretation of the inner integration as an additional Fourier 
transformation
and decomposition of the sum into single terms $I_\nu(\omega)$ results in:

\be
I_{\nu}(\omega)
=
\frac{e^{-F}}{\nu!}\frac{{\cal F}_{\phi\rightarrow-\omega}}{2\pi}
\left\{
{\cal F}_{\Omega\rightarrow-\phi}
[\eta(\Omega)a_{+1}(\Omega)e^{-\beta\hbar\Omega}]
+
{\cal F}_{\Omega\rightarrow+\phi}
[\eta(\Omega)a_{+1}(\Omega)]
\right\}^\nu,
\ee

where ${\cal F}_{\omega\rightarrow+\phi}$ denotes a Fourier transformation and
where $(2\pi)^{-1}{\cal F}_{\phi\rightarrow-\omega}$ is its inverse. The
evaluation of the first two terms is independent of the form of the density of 
modes or the shift amplitude:

\be
I_0(\omega)&=&e^{-F}\delta(\omega)
\nn
I_1(\omega)
&=&
e^{-F}\left\{
\eta(-\omega)a_{+1}(-\omega)e^{+\beta\hbar\omega}
+
\eta(+\omega)a_{+1}(+\omega)\right\}.
\ee

Terms with $\nu\ge2$ could be reexpressed with the help of the binomial theorem
and by noting that products become convolutions after a Fourier 
transformation. Then one sees that the term with index $\nu$ contains the 
effect of events on the shift spectrum during which $\nu$ phonons are 
converted, i.e., created or annihilated.


\section{Application}

Now, we would like to evaluate explicitely the shift spectrum $I(\omega)$ in
harmonic approximation for the isotropic Debye model. 
The Debye model is based on the assumption that the crystal is a 
$d$-dimensional hypercubic lattice. So, it is a Bravais-crystal in which no 
optical modes exist. Isotropy leads to degenerate modes on all $d'=d$ 
branches. The frequencies follow a linear dispersion relation: 
$\Omega(\vec k)=sk$ with the speed of sound $s$ and the absolute value $k$ of 
the wave-number vector $\vec k$. Additionally, the integration over the first 
Brillouin zone is replaced by an integration over the volume of a sphere in 
$k$-space. The size of the sphere is chosen in such a way that it contains $N$ 
allowed wave-number vectors where $N$ is the number of ions in the crystal.
Evaluation of equation (\ref{densityofmodes}) under these assumptions leads to:

\be
\eta(\Omega)
=
d^2\frac{N}{\Omega_D}\left(\frac{\Omega}{\Omega_D}\right)^{d-1}
\theta(\Omega_D-\Omega)\theta(\Omega),
\ee

with the Debye frequency $\Omega_D=sk_D$. As mentioned above, the Debye
frequency replaces the maximal frequency $\Omega_M$. The transition 
probabilities which are the thermally averaged absolute squares of the 
transition matrix-elements $\left<\{n_J'+m_J'\}|e^{iqx}|\{n_J'\}\right>$ 
are to be evaluated in a harmonic system. Here, $q$ stands for the momentum 
transfer and $x$ for the spatial coordinate along its direction. 
After some calculations one finds:

\be
A_{m_J}^J
=
e^{+m_J\beta\hbar\Omega_J/2}
\exp\left\{-\rho^J\coth\left(\frac{\beta\hbar\Omega_J}{2}\right)\right\}
{\rm I}_{m_J}
\left\{\rho^J\left[
\sinh\left(\frac{\beta\hbar\Omega_J}{2}\right)\right
]^{-1}\right\}
=
a_{m_J}^J+\delta_{J,0},
\nn
\label{transitionamplitude}
\ee

with the parameter:

\be
\rho^J=\frac{1}{N}\frac{(\hbar q)^2}{2M_{ion}\hbar\Omega_J},
\ee

and where ${\rm I}_m$ stands for the modified Bessel function of the first 
kind and $M_{ion}$ for the mass of the ion whose nucleus absorbs the 
$\gamma$-photon. For small arguments, we could use the lowest order 
approximation to the Bessel functions:
${\rm I}_m(w)=(w/2)^m/m!+{\cal O}(w^{m+2})$. 
However, the argument of the 
Bessel function is divergent for a vanishing circular frequency 
$\Omega$. Due to the periodic boundary conditions there exists a minimal 
circular frequency $\Omega_1$ which, for a constant volume density of atoms 
$n$, depends on the total number of atoms according to: 
$\Omega_1=\Xi N^{-1/d}$. The constant
of proportionality $\Xi$ depends on the geometry of the lattice and for a 
$d$-dimensional hypercubic lattice is given by $\Xi=2\pi sn^{-1/d}$. For 
large $N$, the argument of the Bessel function becomes: 
$w=2q^2N^{(2-d)/d}/(M_{ion}\beta\Xi^2)$. In three dimensions it 
goes to zero as $N$ goes to infinity. In two dimensions it is independent of 
the number of atoms $N$, so for the validity of the lowest order approximation 
we have to postulate additionally: $2q^2/(M_{ion}\beta\Xi^2)<<1$. In one 
dimension the argument $w$ is not boundend from above and so the lowest order 
approximation is not adequate. In every case where the lowest order 
approximation is justified, we have:

\be
a_{m_J}^J
=
\frac{e^{+m_J\beta\hbar\Omega_J/2}}{|m_J|!}
\left\{\frac{\rho^J}{2}\left[
\sinh\left(\frac{\beta\hbar\Omega_J}{2}\right)\right
]^{-1}\right\}^{|m_J|}
\sim N^{-|m_J|}
\ee

for $|m_J|\ge 1$ and

\be
a_{0}^J
=
-\rho^J\coth\left(\frac{\beta\hbar\Omega_J}{2}\right)
\sim N^{-1}
\ee

for $m_J=0$. The density of modes $\eta(\Omega)$ only contributes one 
additional factor of $N$. So, if $|m_J|>1$ the negative powers of the number 
of ions in the crystal cannot be compensated and the corresponding 
terms do not contribute. This result has already been used to derive equation
(\ref{pde}). For vanishing absolute temperature, the above expansion is always 
legitimate and one finds:

\be
a_{+1}^J=+\rho^J~~~a_{0}^J=-\rho^J~~~a_{-1}^J=0.
\ee

Finally, the transition to the continuum has to be performed as described 
above. Now, equation (\ref{fouriershift}) can be reexpressed depending on the 
dimension $d$ of the crystal:

\be
\tilde I_d(\phi)
=
\exp\left\{
\frac{\sigma_d}{\Omega_D^{d-1}}
\int_0^{\Omega_D}d\Omega\Omega^{d-2}
\left[
\coth\left\{\frac{\beta\hbar\Omega}{2}\right\}(\cos\{\phi\Omega\}-1)
-
i\sin\{\phi\Omega\}
\right]\right\}
\ee

with the parameter: $\sigma_d=(dq\hbar)^2/(2M_{ion}\hbar\Omega_D)$. The 
aforementioned
separation into a $\phi$-dependent and a $\phi$-independent part without 
leaving a non-integrable pole in the $\phi$-dependent part is possible in three 
dimensions at any temperature and in two dimensions for zero temperature. 
The analytic expression for the exponent $F_d$ of the 
M\"ossbauer-Lamb fraction is given by:

\be
F_d
=
\frac{\sigma_d}{\Omega_D^{d-1}}
\int_0^{\Omega_D}d\Omega\Omega^{d-2}
\coth\left\{\frac{\beta\hbar\Omega}{2}\right\}
\ee

which for zero temperature is equal to ${F_d^0}(\Omega_D)=\sigma_d/(d-1)$ 
and for non-zero temperature and in three dimensions reads:

\be
F_3
=
\sigma_3\left[
\frac{\ln(1-e^{+\beta\hbar\Omega_D})}{\beta\hbar\Omega_D}
-
\frac
{{\rm dilog}(e^{+\beta\hbar\Omega_D})+{\rm dilog}(1-e^{+\beta\hbar\Omega_D})}
{(\beta\hbar\Omega_D)^2}
+
\frac{\pi^2}{6(\beta\hbar\Omega_D)^2}
\right]
\ee

where the dilogarithm is defined as in chapter 27.7 of reference 
\cite{abramowitzstegun}. For the 
subset of cases investigated up to the present, the exponent is bounded and 
the Fourier retransformation can be carried out term by term for the  
exponential series. The integrated contributions $\tilde I_{d,\nu}(\phi=0)$
follow a Poissonian distribution with a mean value equal to $F_d$. 
So, this parameter determines how fast the series converges. For zero 
temperature the retransformations for all the terms of the series can be 
carried out exactly:

\be
I_{3,\nu}^0(\omega)
=
\frac{e^{-\sigma_3/2}}{\Omega_D}
\sum_{\mu_1=0}^\nu\sum_{\mu_2=0}^{\mu_1}
\frac{(-1)^{\mu_1}{{\sigma_3}^\nu}}{\nu!(2\nu-\mu_2-1)!}
\binom{\nu}{\mu_1}
\binom{\mu_1}{\mu_2}
\theta\left(\frac{\omega}{\Omega_D}-\mu_1\right)
\left(\frac{\omega}{\Omega_D}-\mu_1\right)^{2\nu-\mu_2-1}
\ee

(see figure \ref{d3t0}) and

\be
I_{2,\nu}^0(\omega)
=
\frac{e^{-\sigma_2}}{\Omega_D}
\sum_{\mu=0}^\nu
\frac{(-1)^{\mu}{{\sigma_2}^\nu}}{\nu!(\nu-1)!}
\binom{\nu}{\mu}
\theta\left(\frac{\omega}{\Omega_D}-\mu\right)
\left(\frac{\omega}{\Omega_D}-\mu\right)^{\nu-1}
\ee

(see figure \ref{d2t0}).
The integrations in the three dimensional case and at finite temperatures can 
be carried out after approximating the hyperbolic cotangent by:

\be
\coth\left\{\frac{\beta\hbar\Omega}{2}\right\}
\approx
1+2\frac{e^{-\beta\hbar\Omega/2}}{\beta\hbar\Omega}.
\label{approximation}
\ee

An expansion into partial fractions leads to (see figure \ref{d3t}):

\be
I_{3,\nu}(\omega)
\approx
\sum_{\mu_1=0}^\nu\sum_{\mu_2=0}^{\mu_1}
\sum_{\lambda_1=1}^\nu
&\Lambda'_{1,\lambda_1}&
(-\omega')^{\lambda_1-1}
e^{+\beta\hbar\Omega_D\omega'}
\theta(-\omega')
+
\nn
+
\sum_{\lambda_2=1}^{2(\nu-\mu_2)}
&\Lambda'_{2,\lambda_2}&
(+\omega')^{\lambda_2-1}
\theta(+\omega')
+
\nn
+
\sum_{\lambda_3=1}^{\nu-\mu_2}
&\Lambda'_{3,\lambda_3}&
(+\omega')^{\lambda_3-1}
e^{-\beta\hbar\Omega_D\omega'}
\theta(+\omega')
\ee

with the factors independent of $\omega'=2\mu_2-\mu_1+\omega/\Omega_D$ 
defined as:

\be
\Lambda'_{(1,2,3),\lambda}
=
e^{-F}
\frac{1}{\Omega_D}
\frac{{\sigma_3}^\nu}{\nu!}
\left(\frac{\beta\hbar\Omega_D}{2}\right)^{2\nu-\mu_1}
\binom{\nu}{\mu_1}
\binom{\mu_1}{\mu_2}
e^{-\mu_2\beta\hbar\Omega/2}
\frac{\Lambda_{(1,2,3),\lambda}}{{\Omega_D}^{4\nu-\mu_1-\lambda}}
\frac{(-1)^\lambda}{(\lambda-1)!}
\ee

and where the coefficients $\Lambda_{(1,2,3),\lambda}$ are given by:

\be
\Lambda_{1,\lambda_1}
&=&
\frac{1}{(\lambda_1-1)!}
\left.\frac{d^{\nu-\lambda_1}}{d\phi^{\nu-\lambda_1}}\right|_{i\phi=+\beta\hbar}
\frac
{p^{\mu_1-\mu_2}(\phi)}
{(-i\phi)^{2(\nu-\mu_2)}(-i\phi-\beta\hbar)^{\nu-\mu_2}}
\nn
\Lambda_{2,\lambda_2}
&=&
\frac{1}{(\lambda_2-1)!}
\left.\frac{d^{2(\nu-\mu_2)-\lambda_2}}{d\phi^{2(\nu-\mu_2)-\lambda_2}}
\right|_{\phi=0}
\frac
{p^{\mu_1-\mu_2}(\phi)}
{(-i\phi-\beta\hbar)^{\nu-\mu_2}(+i\phi-\beta\hbar)^\nu}
\nn
\Lambda_{3,\lambda_3}
&=&
\frac{1}{(\lambda_3-1)!}
\left.\frac{d^{\nu-\mu_2-\lambda_3}}{d\phi^{\nu-\mu_2-\lambda_3}}
\right|_{i\phi=-\beta\hbar}
\frac
{p^{\mu_1-\mu_2}(\phi)}
{(-i\phi)^{2(\nu-\mu_2)}(+i\phi-\beta\hbar)^\nu}
\ee

with the polynomial in $\phi$:

\be
p(\phi)
=
-i\phi^3(\beta\hbar\Omega_D+e^{-\beta\hbar\Omega_D/2})
-\phi^2\beta\hbar(1-e^{-\beta\hbar\Omega_D/2}/2)
-i\phi(\beta\hbar)^2(\beta\hbar\Omega_D/4)
-(\beta\hbar)^3/4
\nn
\ee

For the two-dimensional case at finite temperatures, the separation into a 
$\phi$-dependent and a $\phi$-independent part is no longer possible without 
leaving behind a non-integrable pole. Actually, the M\"ossbauer-Lamb fraction
$e^{-F}$ is exactly equal to zero because its exponent diverges.
Hence, there is no recoil-free emission in this case. For further 
evaluation, we start by separating off the shift spectrum for zero temperature 
from that for finite temperature:

\be
\tilde I_2(\phi)
=
\tilde I_2^0(\phi)
\exp\left\{
2\frac{\sigma_2}{\Omega_D}
\int_0^{\Omega_D}d\Omega
\frac{e^{-\beta\hbar\Omega}}{1-e^{-\beta\hbar\Omega}}
(\cos\{\phi\Omega\}-1)
\right\}
\ee

With an approximation analogous to that in equation (\ref{approximation}) 
one finds:

\be
\tilde I_2(\phi)
\approx
{\tilde I^0_2(\phi)}
\frac{(\beta\hbar\Omega_D/2)^2}{(\beta\hbar\Omega_D/2)^2+(\phi\Omega_D)^2}
\exp\left\{
2\frac{\sigma_2}{\beta\hbar\Omega_D}
[E_1(\beta\hbar\Omega_D/2)-{\rm Re}\{E_1[(\beta\hbar/2\pm i\phi)\Omega_D]\}]
\right\},
\nn
\ee

where the exponential integral function $E_1$ is defined as in chapter 5 of 
reference \cite{abramowitzstegun}:

\be
E_1(z)=\int_z^\infty dt\frac{e^{-t}}{t}.
\ee

The following integrations cannot be 
performed analytically, due to the exponential integral function $E_1$.
For the two dimensional case we had already postulated that 
$\sigma_2$ is to be small compared to one. In the physically interesting 
case of absolute temperatures $T$ small compared to the Debye temperature
$\Theta_D=\hbar\Omega_D/k_B$, 
the exponent as such becomes extremely small and the exponential function can 
be approximated by one. In the low-temperature approximation it remains to 
perform the Fourier retransformation of the power of the Lorentz distribution.
However, the asymptotic form of this function for large $\phi$ is given 
by $[\beta\hbar/(2\phi)]^{-2\sigma_2/(\beta\hbar\Omega_D)}$. Due to the small 
absolute values of the exponent this function decays slower than $\phi^{-1}$ 
which renders the Fourier retransformation difficult. Let us introduce an 
additional factor which ensures the convergence of the integral under 
investigation. If this manoeuvre is performed in a suitable way, the 
observable spectrum $S(\omega_\gamma)$ can be obtained directly. When 
choosing a Lorentz distribution as a model for the natural line shape of the 
nucleus, the observabel spectrum is just the convolution of the former with 
the shift spectrum $I(\omega)$ in momentum representation. This corresponds 
to a multiplication of the Fourier transformed shift spectrum $\tilde I(\phi)$ 
with the Fourier transformed Lorentz distribution, i.e., a Laplace 
distribution also known as a doubly-exponential distribution. What remains to 
be executed is:

\be
S_2(\omega)
\approx
S_0~{I^0_2(\omega)}
*
\int\frac{d\phi}{2\pi}e^{i\phi(\omega-\omega_0)}e^{-\zeta|\phi|}
\left(
\frac
{(\beta\hbar\Omega_D/2)^2}
{(\beta\hbar\Omega_D/2)^2+(\phi\Omega_D)^2}
\right)^{\frac{\sigma_2}{\beta\hbar\Omega_D}},
\ee

with the resonance frequency of the nucleus $\omega_0$, the natural line-width 
$\zeta$ (half-width half-maximum) and the integrated intensity $S_0$. The 
integral can be interpreted as a Laplace transformation and yields 
(see \cite{erdelyi} and figure \ref{d2tc}):

\be
S_2(\omega)
\approx
\frac{\beta\hbar S_0}{2\pi}{I_2^0}(\omega)*{\rm Re}\left\{
{\bf\rm L}_{\frac{1}{2}-\frac{\sigma_2}{\beta\hbar\Omega_D}}
\left(\frac{\beta\hbar[\zeta-i(\omega-\omega_0)]}{2}\right)\right\},
\label{line}
\ee

with:

\be
{\bf\rm L}_l(z)=2^{l-1}\sqrt{\pi}{\rm\Gamma}\left(l+\frac{1}{2}\right)z^{-l}
[{\bf\rm H}_l(z)-{\rm Y}_l(z)],
\ee

where ${\bf\rm H}_l(z)$ denotes the Struve function (see chapter 12 of 
reference \cite{abramowitzstegun}), ${\rm Y}_l(z)$ stands for a Bessel 
function of the second kind also known as Weber function (see chapter 9 of
reference \cite{abramowitzstegun}), and $\Gamma(z)$ represents the 
$\Gamma$-function. In the limit of vanishing line width $\zeta$, the 
real part of the argument of ${\bf\rm L}_l(z)$ is equal to zero. Now, the 
approximation to what shall be called the {\it Quasi-M\"ossbauer line} of the 
shift spectrum is the convolution of the zeroth order term of the shift 
spectrum for zero temperature $I_{2,\nu=0}^0(\omega)$ with the line shape given by 
equation (\ref{line}) with $S_0$, $\omega_0$, and $\zeta$ set to zero (see
figure \ref{d2tab}):

\be
I_{2,\nu=0}^T(\omega)
\approx
\frac{e^{-\sigma_2}}{\Omega_D}
\frac{\beta\hbar\Omega_D}{2\pi}
{\rm Re}\left\{
{\bf\rm L}_{\frac{1}{2}-\frac{\sigma_2}{\beta\hbar\Omega_D}}
\left(\frac{\beta\hbar[\zeta-i(\omega-\omega_0)]}{2}\right)\right\}.
\ee


\section{Results}

In three dimensions there is largely agreement between the present work and 
\cite{visscher}. Here, a deeper insight has been permitted due to the 
presentation of the results in an analytic form. The investigations have been 
extended to incorporate the 
two dimensional case at zero and at finite temperature respectively. The one-dimensional case
at zero temperature is also covered but not investigated any further here.

In the two dimensional case at finite absolute temperatures, the 
M\"ossbauer-Lamb fraction according to its definition (\ref{mlf}) 
vanishes exactly. 
No recoil-free absorption or emission takes place and strictly speaking, 
there is no M\"ossbauer line in the spectra. The mathematical investigation 
carried out here in this case is only applicable for small values of the 
model parameter $\sigma_2$. But even then the execution of the calculations
was not possible in the same way as for the other cases (three dimensions 
and/or zero absolute temperature). An alternative way had to be pursued.
Although a M\"ossbauer line does not exist, a central peak is 
found. Already in the shift-spectrum representation this Quasi-M\"ossbauer 
peak shows a finite width. The width depends on the model parameter 
$\sigma_2$ and the temperature $T$. 

The disappearance of the M\"ossbauer-Lamb fraction in two dimensions and at
finite absolute temperature signifies that no entirely recoil-free events are 
possible and that there cannot be a M\"ossbauer peak. Its replacement, the 
Quasi-M\"ossbauer line must be based on quasi-recoil-free processes. These are
events during which only very little energy is transformed into lattice 
vibrations. This could be due to the predominant creation and annihilation of
low-energy phonons. Another more general possibility is based on a dominance of 
multi-phonon events with small net energy. In the two-dimensional case, as 
opposed to the three-dimensional, the density of modes does not drop rapidly 
enough for small absolute values of the circular frequency in order to 
compensate the pole of the shift amplitudes. This indicates that the first of 
the two explanations given above is the decisive one but does not rule out the
second. The different mechanism of formation in two dimensions leads to a 
central peak of finite width depending on the temperature. 
In the representation as observable spectrum, i.e., after the
convolution of the shift spectrum with the natural line shape, the
M\"ossbauer and the Quasi-M\"ossbauer line appear qualitatively similar (see
figure \ref{d2tab}). Quantitatively they are not. They differ in peak
height, line width, and integrated intensity. Given the qualitative
similarity, the calculations would have to be reperformed with a more
realistic than the Debye model in order to obtain more reliable quantitative
results.

While the phonon wings might render the determination of the central line's 
integrated intensity cumbersome, they provide the best means to identify
two-dimensional elastic behavior (see figure \ref{d2tc}). For narrow spectra,
in three dimensions,
the phonon wing passes through a relative minimum almost reaching down to
zero close to the central peak. In two dimensions a nearly horizontal
passage can be seen. This means also that the phonon wings for positive and
negative circular frequencies are continuously connected at zero frequency
in three dimensions (see also figure \ref{d3t}). In two dimensions there is
a step (see figure \ref{d2tc}).
 
Room for extension of the present work lies in the investigation of slowly 
decaying spectra in two dimensions.


\section*{Acknowledgements}

Help by R.~J.~Jelitto is gratefully acknowledged.


\figu{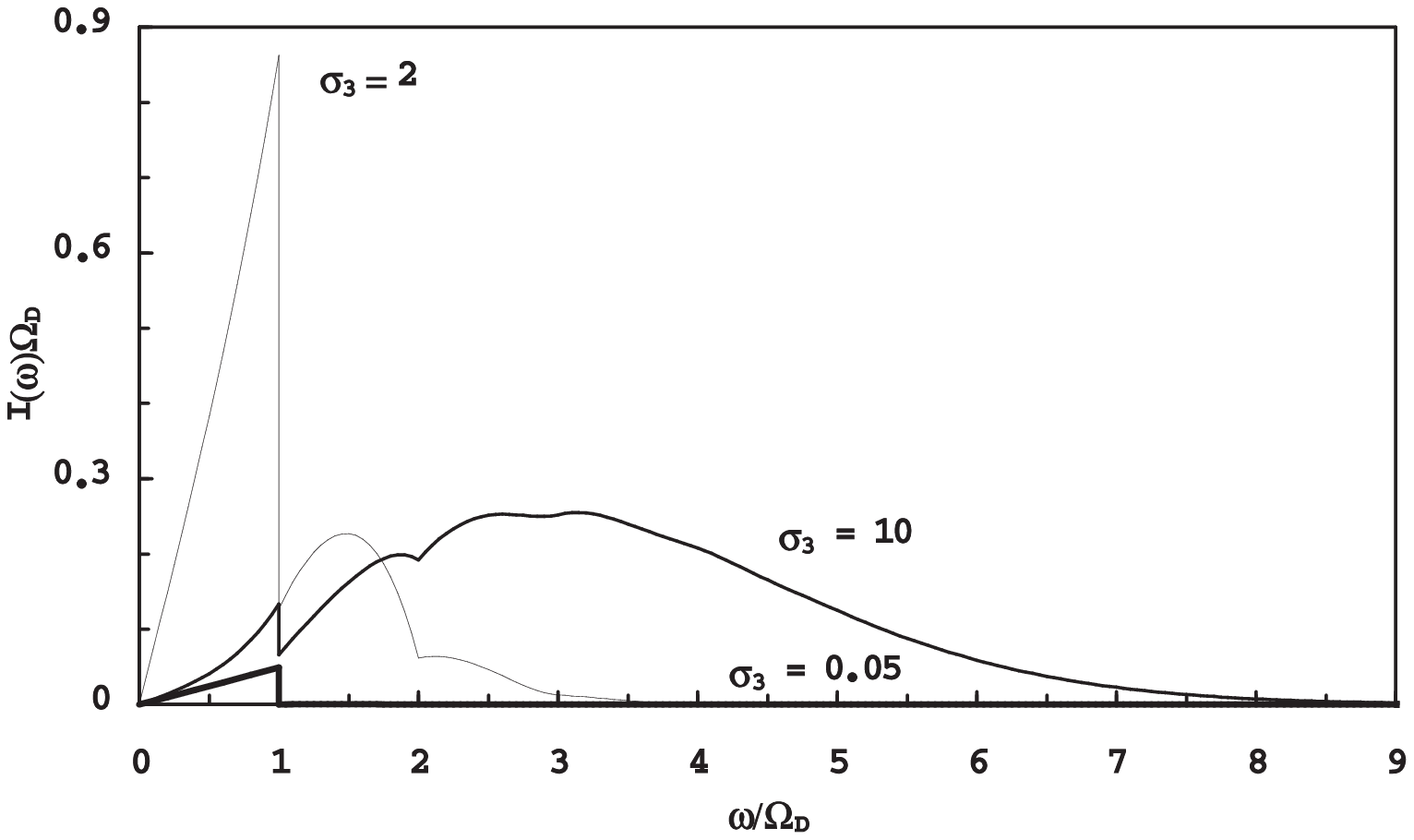}{
The figure shows the shift spectrum $I(\omega)$ in three dimensions and for
zero temperature. Different values of the model parameter $\sigma_3$ have 
been chosen. The abscissa and the ordinate are rescaled with the Debye
frequency $\Omega_D$. In an unscaled presentation, a variation of the Debye 
frequency $\Omega_D$ with the parameter $\sigma_3$ kept constant would lead
to variations of the aspect ratio of the graph. The step structure inherent to 
the shift spectrum with steps at integer multiples of the Debye frequency 
emerges from the plot. For smaller values of the M\"ossbauer-Lamb fraction 
these steps are smoothed out more and more until the spectrum resembles a 
Poissonian or Gaussian distribution. For large M\"ossbauer-Lamb fractions apart 
from the M\"ossbauer line which coincides with the axis in this plot, the 
spectrum mainly only consists of a small contribution between zero and the 
Debye frequency $\Omega_D$. At zero temperature, the shift spectrum is 
identical to zero for negative values of the frequency $\omega$, because there 
is no initial population of the crystal with phonons which could be 
annihilated. Especially the sharp peak at $\omega=\Omega_D$ can be 
traced 
back to the use of the Debye model and would be less pronounced for a model
with an adapted behavior for higher phonon frequencies.}
\figu{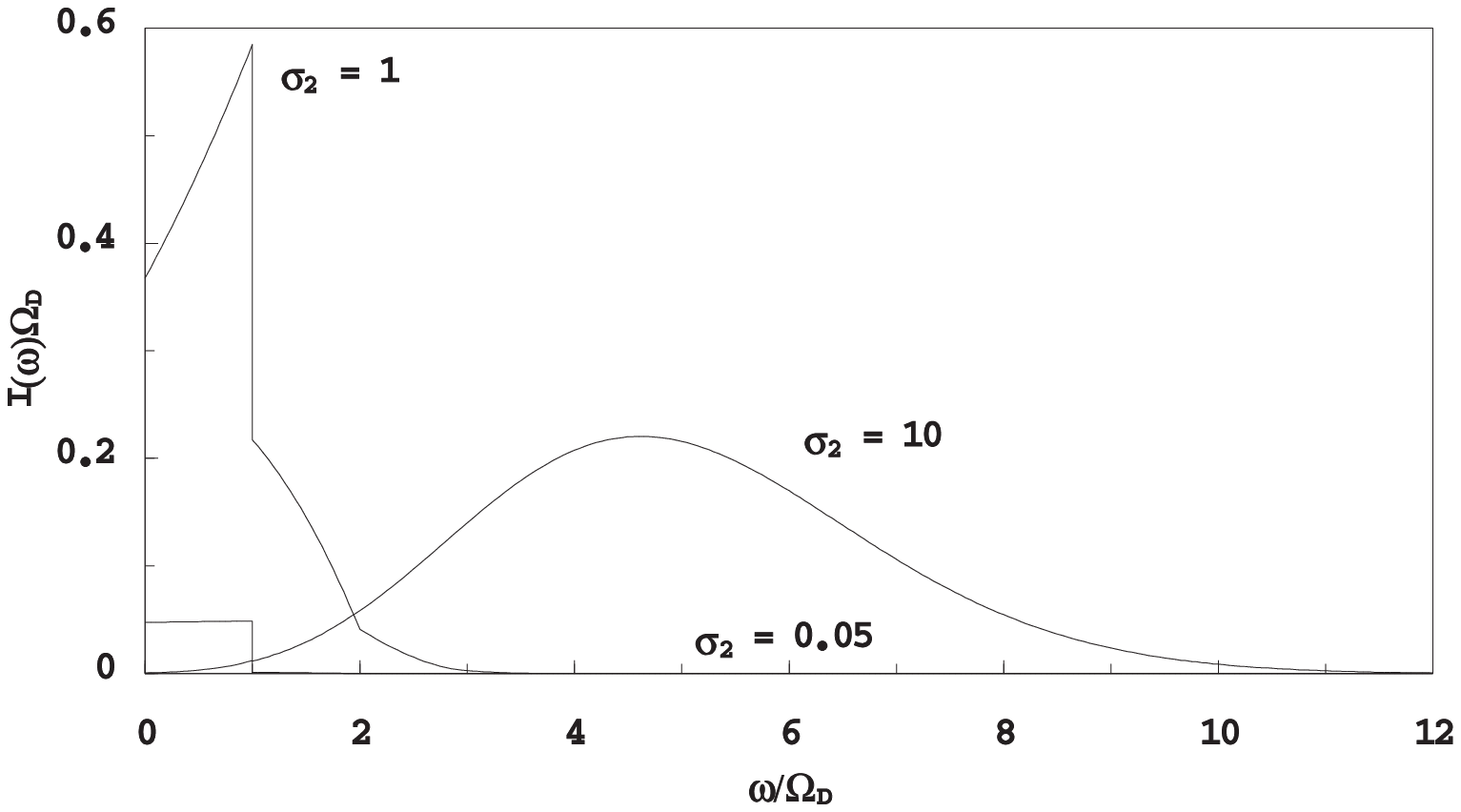}{
Here, the shift spectrum $I(\omega)$ for two dimensions is depicted for
zero temperature, where various values have been chosen for the model 
parameter $\sigma_2$. The scaling properties and the step structure are equally well 
discernible as in figure \ref{d3t0}. The transition to a Poissonian or Gaussian like
distribution for slowly decaying spectra becomes evident, too. The 
difference to the three-dimensional case lies in the details of the different
contributions. The spectrum for the highest M\"ossbauer-Lamb fraction in two
dimensions shows an almost constant passage from zero frequency to the Debye
frequency $\Omega_D$ whereas in three dimensions it shows a linearly rising 
behaviour.
}
\figu{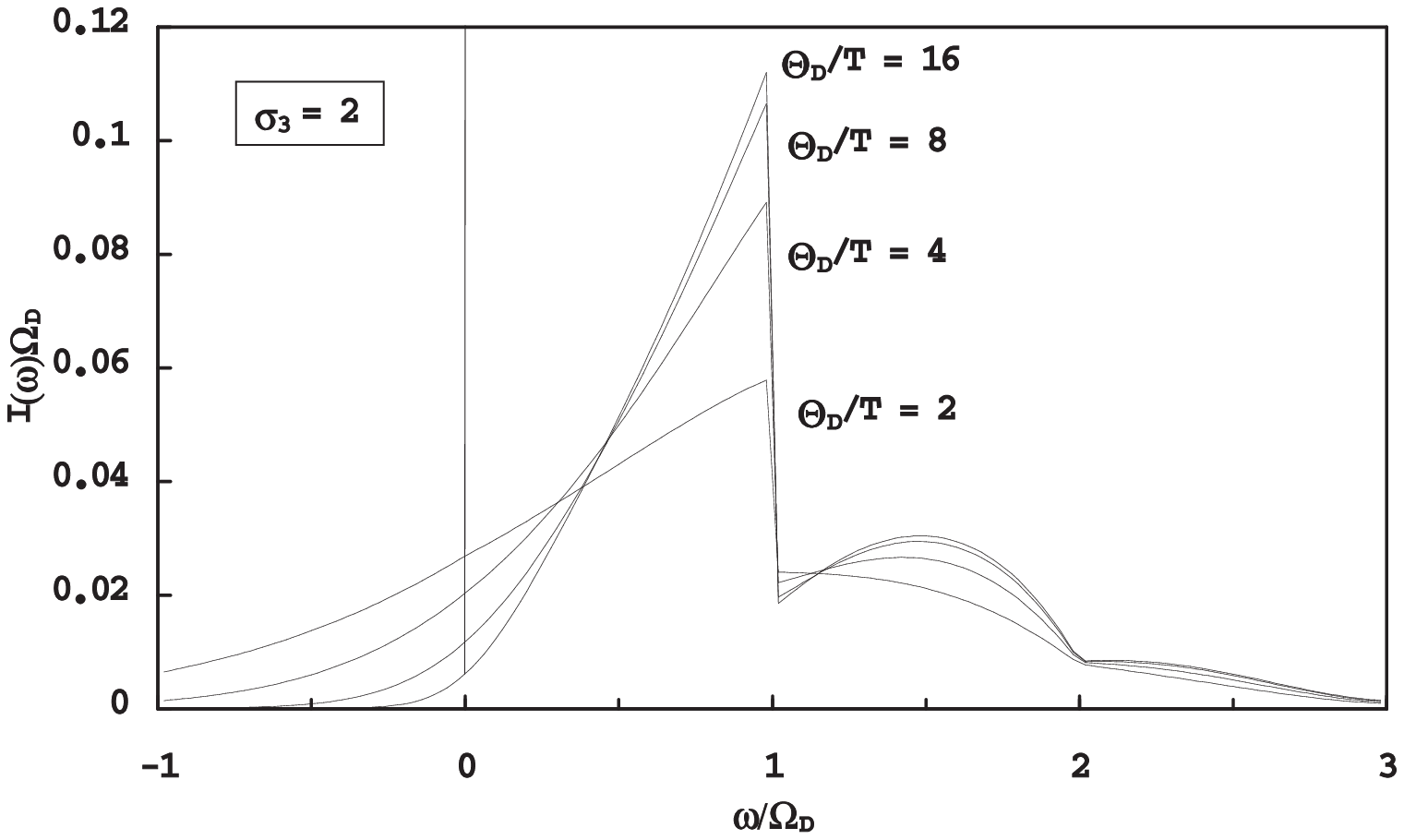}{
In this figure, the shift spectrum $I(\omega)$ is presented for different 
values of the absolute temperature $T$ and the model parameter:
$\sigma_3=2$. For zero absolute temperature the corresponding plot is given 
in figure \ref{d3t0}. The shift spectrum only depends on the ratio of the absolute 
temperature to the Debye temperature $\Theta_D$. For the shown cases with 
relatively low temperatures, the step structure known from figure \ref{d3t0} is 
conserved. However, for increasing temperature it becomes smoother. The 
scaling behaviour with the Debye frequency $\Omega_D$ also remains the same, 
only that additionally, the ratio between the absolute and the Debye 
temperature $\Theta_D$ has to be kept constant. Now there are also
contributions 
for negative circular frequencies $\omega$. They belong to those events where 
net energy is drawn from the crystal.
}
\figu{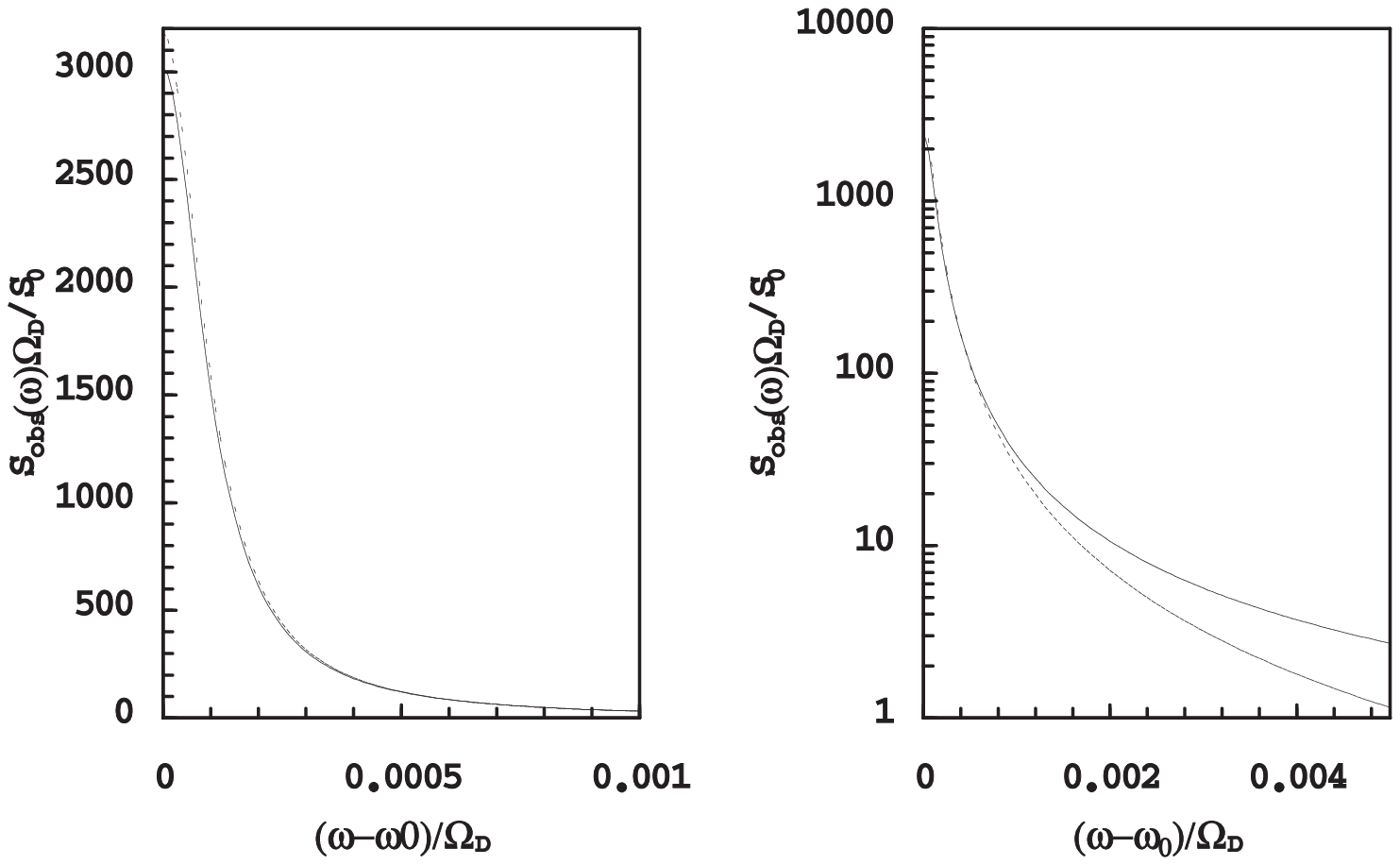}{
The two figures \ref{d2tab} serve as a comparison between the 
Quasi-M\"ossbauer
line in two dimensions and the M\"ossbauer line in three dimensions for the 
same choice of the physical constants: $\sigma_3=0.05=(9/4)\sigma_2$, 
$\beta\hbar\Omega_D=10$, and $\zeta/\Omega_D=10^{-4}$. The two graphs show the
observable spectrum $S_{obs}(\omega)$ rescaled with the Debye frequency 
$\Omega_D$ and the integrated intensity $S_0$. In
order to emphasise the differences, a small natural line-width has been chosen.
The lines are always even functions with respect to inversion of the sign of 
the argument relative to the resonance frequency $\omega_0$. The figure on
the lhs is a doubly linear plot. Here, the slightly higher maximum value of the M\"ossbauer
line (dashed) compared to the Quasi-M\"ossbauer line (solid) can be seen. 
For large values of the circular frequency, the M\"ossbauer line tends to zero 
faster than the Quasi-M\"ossbauer line. This can best be seen in the figure
on the rhs 
which has a logarithmic ordinate. The two lines do not have the same 
integrated area. The area beneath the M\"ossbauer line is larger by a factor 
of $e^{(7/4)\sigma_3}$. On first sight, the two lines do not differ in their 
width. Numerically a difference of a Quasi-M\"ossbauer line with a 0.4 percent 
larger width is found. This value can be increased to 3 percent for the 
smallest natural line width found in nature ($\zeta/\Omega_D=10^{-11}$). In 
the last case however, the absolute difference is one million times smaller.
}

\figu{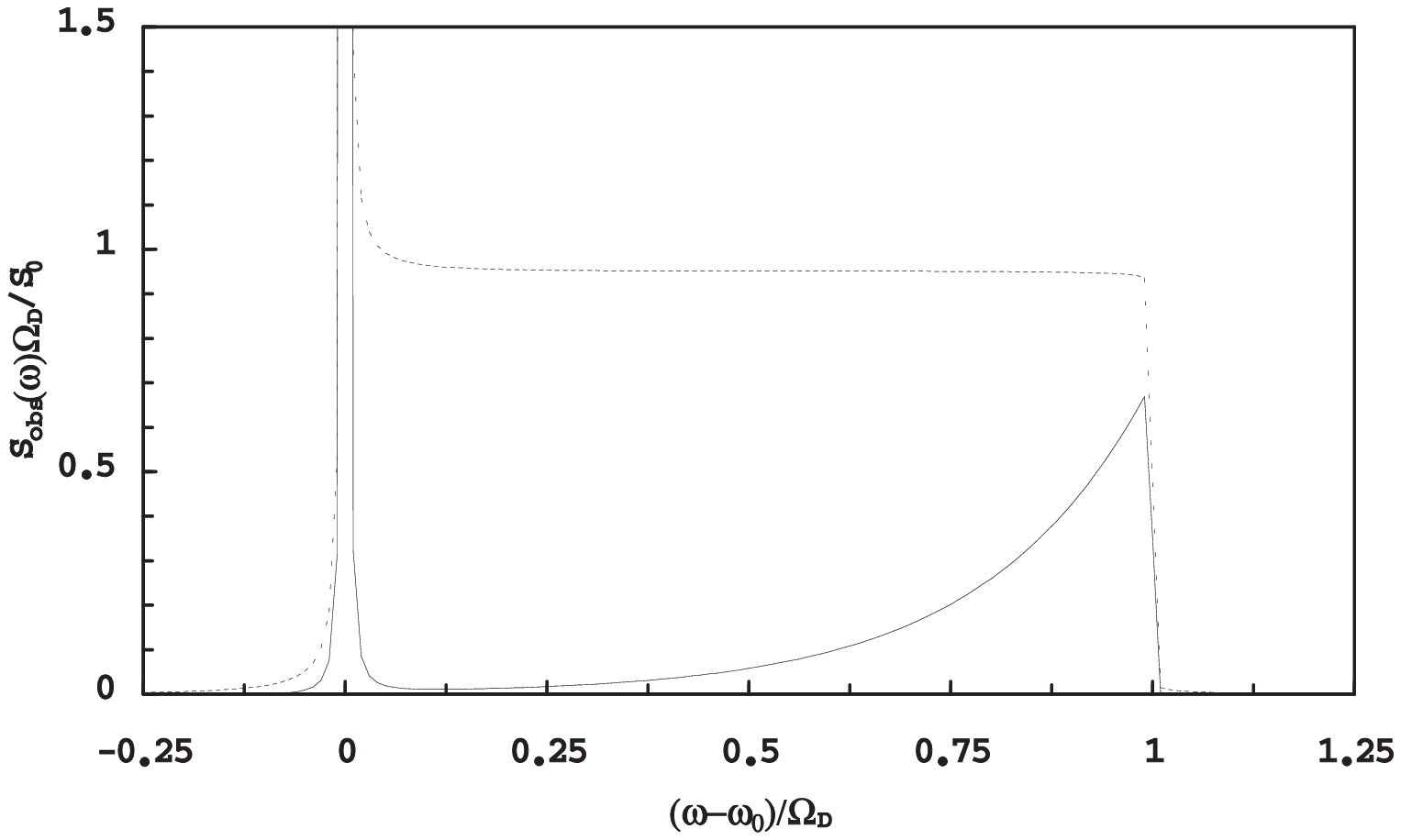}{
In this figure, the graphs of the following two observable spectra 
$S_{obs}(\omega)$
at finite temperature are depicted: The one for two dimensions (dashed) with 
the model parameter set to $\sigma_2=1/45$ and the other for three 
dimensions (solid) with the model parameter $\sigma_3=1/20$. The other 
parameters defining the temperature and the natural line-width respectively
are: 
$\beta\hbar\Omega_D=10$ and $\zeta/\Omega_D=10^{-4}$. The spectrum in two 
dimensions decays more slowly for negative values of the circular frequency 
$\omega$ than in three dimensions. The most pronounced difference can be found for 
the circular frequency between zero and the Debye frequency $\Omega_D$.
There,
the contribution of the first order of the expansion of the exponential 
function shows up. In three dimensions, the observable spectrum 
$S_{obs}(\omega)$ passes through a relative minimum almost reaching down to 
zero. In two dimensions, a nearly horizontal passage can be seen. This 
interval is the most suitable place to distinguish between the two cases by 
measurements and to identify two-dimensional elastic behaviour.
}


\end{document}